\documentclass[a4paper,12pt]{article}
\pdfoutput=1
\usepackage{feynmp-auto,expdlist}
\usepackage{amsmath, amsfonts}
\usepackage{graphicx,arydshln}
\usepackage{enumerate}
\usepackage{hyperref}
\usepackage{latexsym}
\usepackage{dsfont}
\usepackage{hepnicenames}
\usepackage{enumerate}
\usepackage{soul}
\usepackage[normalem]{ulem}
\usepackage{comment}

\newcommand{\footnoten}[1]{}
\newcommand\mpl{M_{\rm Pl}}
\usepackage[font={small},textfont={it}]{caption} 

\usepackage{units}

\newcommand{\Mtexp}{172.6}

\newcommand{\Mhexp}{125.1}

\newcommand{\Mtdiff}{ \bigg(\frac{M_t}{\GeV}-\Mtexp\bigg)}
\newcommand{\asdiff}{\, \frac{\alpha_3(M_Z)-0.1179}{0.0009} }
\newcommand{\Mhdiff}{\bigg(\frac{M_h}{\GeV}-\Mhexp\bigg)}

\newcommand{\va}{{v}}
\newcommand{\wa}{{w}}
\newcommand{\vecf}[1]{#1}

\usepackage{mathrsfs,graphicx,rotating,amsmath,amsfonts,mathtools,booktabs,amssymb,wasysym}
\usepackage{hyperref}\usepackage{slashed}
\usepackage[nosort]{cite}
\usepackage[table,xcdraw,dvipsnames]{xcolor}
\usepackage{bm}
\usepackage{multirow,multicol}
\hypersetup{colorlinks,bookmarksopen,bookmarksnumbered,
	linkcolor=blus,pdfstartview=FitH,urlcolor=rossos,citecolor=verde}
\allowdisplaybreaks

\newcommand{\km}{\,{\rm km}}
\renewcommand{\[}{\left[}

\def\Lag{\mathscr{L}}

\newcommand{\mio}[1]{}

\newcommand{\med}[1]{\langle #1\rangle}

\def\bpm{\begin{pmatrix}}
	\def\epm{\end{pmatrix}}

\usepackage{mathrsfs}

\newcommand{\fig}[1]{~\ref{fig:#1}}
\newcommand{\sfrac}[2]{#1/#2}

\allowdisplaybreaks
\usepackage{multicol}
\usepackage{color}
\definecolor{rosso}{cmyk}{0,1,1,0.4}
\definecolor{rossos}{cmyk}{0,1,1,0.55}
\definecolor{rossoc}{cmyk}{0,1,1,0.2}
\definecolor{blu}{cmyk}{1,1,0,0.3}
\definecolor{blus}{cmyk}{1,1,0,0.6}
\definecolor{bluc}{cmyk}{1,1,0,0.1}
\definecolor{verde}{cmyk}{0.92,0,0.59,0.25}
\definecolor{verdec}{cmyk}{0.92,0,0.59,0.15}
\definecolor{verdes}{cmyk}{0.92,0,0.59,0.4}

\newcommand{\eq}[1]{~{\rm (\ref{eq:#1})}}

\newcommand{\MeV}{\,{\rm MeV}}
\newcommand{\GeV}{\,{\rm GeV}}
\newcommand{\TeV}{\,{\rm TeV}}

\def\circa#1{\,\raise.3ex\hbox{$#1$\kern-.75em\lower1ex\hbox{$\sim$}}\,}

\newcommand{\beq}{\begin{equation}}
\newcommand{\eeq}{\end{equation}}

\newcommand{\bea}{\begin{eqnarray}}
\newcommand{\eea}{\end{eqnarray}}
\newcommand{\be}{\begin{equation}}
\newcommand{\ee}{\end{equation}}
\font\tenrsfs=rsfs10 at 12pt
\font\sevenrsfs=rsfs7
\font\fiversfs=rsfs5
\newfam\rsfsfam
\textfont\rsfsfam=\tenrsfs
\scriptfont\rsfsfam=\sevenrsfs
\scriptscriptfont\rsfsfam=\fiversfs

\newsavebox\MBox

\renewenvironment{thebibliography}[1]
{\begin{multicols}{2}[\section*{\refname}]%
		\@mkboth{\MakeUppercase\refname}{\MakeUppercase\refname}%
		\list{\@biblabel{\@arabic\c@enumiv}}%
		{\settowidth\labelwidth{\@biblabel{#1}}%
			\leftmargin\labelwidth
			\advance\leftmargin\labelsep
			\@openbib@code
			\usecounter{enumiv}%
			\let\p@enumiv\@empty
			\renewcommand\theenumiv{\@arabic\c@enumiv}}%
		\sloppy
		\clubpenalty4000
		\@clubpenalty \clubpenalty
		\widowpenalty4000%
		\sfcode`\.\@m}
	{\def\@noitemerr
		{\@latex@warning{Empty `thebibliography' environment}}%
		\endlist\end{multicols}}

\renewcommand{\L}\Lag

\def\circa#1{\,\raise.3ex\hbox{$#1$\kern-.75em\lower1ex\hbox{$\sim$}}\,}
\makeatletter

\font\ital=cmu10

\def\hhref#1{\href{http://arxiv.org/abs/#1}{arXiv:#1}}
\usepackage{xstring}
\newcommand{\hhrefq}[1]{\IfSubStr{#1}{:}{\href{http://inspirehep.net/search?ln=en&ln=en&p=#1&of=hb&action_search=Search&sf=&so=d&rm=&rg=25&sc=0}{InSpire:#1}}{\hhref{#1}}}

\def\art{\@ifnextchar[{\eart}{\oart}}
\def\eart[#1]#2#3#4#5#6{{\rm #2}, {\em #3 \bf #4} {\rm (#6) #5} ({\em #1})}
\def\article{\@ifnextchar[{\earticle}{\oarticle}}
\def\oarticle#1#2#3#4#5#6{{\rm #1}, {\ital `#6'}, {\rm #2 #3 (#5) #4}}
\def\earticle[#1]#2#3#4#5#6#7{{\rm #2}, {\ital `#7'}, {\rm #3 #4 (#6) #5}  [\hhrefq{#1}]}
\def\hepart[#1]#2{{\rm #2, \sl#1}}
\def\heparticle[#1]#2#3{#2, {\ital `#3'} [\hhrefq{#1}]}
\newcommand{\doi}[1]{\href{http://dx.doi.org/#1}{[link]}}

\newcommand{\hhrefqq}[1]{\IfBeginWith{#1}{10.}{\href{https://doi.org/#1}{doi:#1}}{\hhrefq{#1}}}
\def\earticle[#1]#2#3#4#5#6#7{{\rm #2}, {\ital `#7'}, {\rm #3 #4 (#6) #5}  [\hhrefqq{#1}]}

\renewenvironment{thebibliography}[1]
{\begin{multicols}{2}[\section*{\refname}]%
		\@mkboth{\MakeUppercase\refname}{\MakeUppercase\refname}%
		\list{\@biblabel{\@arabic\c@enumiv}}%
		{\settowidth\labelwidth{\@biblabel{#1}}%
			\leftmargin\labelwidth
			\advance\leftmargin\labelsep
			\@openbib@code
			\usecounter{enumiv}%
			\let\p@enumiv\@empty
			\renewcommand\theenumiv{\@arabic\c@enumiv}}%
		\sloppy
		\clubpenalty4000
		\@clubpenalty \clubpenalty
		\widowpenalty4000%
		\sfcode`\.\@m}
	{\def\@noitemerr
		{\@latex@warning{Empty `thebibliography' environment}}%
		\endlist\end{multicols}}

\newcounter{alphaequation}[equation]
\def\thealphaequation{\theequation\hbox to
	0.6em{\hfil\alph{alphaequation}\hfil}}
\def\eqnsystem#1{
	\def\@eqnnum{{\rm (\thealphaequation)}}
	\def\@@eqncr{\let\@tempa\relax \ifcase\@eqcnt \def\@tempa{& & &} \or
		\def\@tempa{& &}\or \def\@tempa{&}\fi\@tempa
		\if@eqnsw\@eqnnum\refstepcounter{alphaequation}\fi
		\global\@eqnswtrue\global\@eqcnt=0\cr}
	\refstepcounter{equation} \let\@currentlabel\theequation \def\@tempb{#1}
	\ifx\@tempb\empty\else\label{#1}\fi
	\refstepcounter{alphaequation}
	\let\@currentlabel\thealphaequation
	\global\@eqnswtrue\global\@eqcnt=0 \tabskip\@centering\let\\=\@eqncr
	$$\halign to \displaywidth\bgroup \@eqnsel\hskip\@centering
	$\displaystyle\tabskip\z@{##}$&\global\@eqcnt\@ne
	\hskip2\arraycolsep\hfil${##}$\hfil& \global\@eqcnt\tw@\hskip2\arraycolsep
	$\displaystyle\tabskip\z@{##}$\hfil
	\tabskip\@centering&\llap{##}\tabskip\z@\cr}
\def\endeqnsystem{\@@eqncr\egroup$$\global\@ignoretrue} \makeatother

\definecolor{Gray}{gray}{0.95}

\def\bal#1\eal{\begin{align}#1\end{align}}

\begin{document}

\thispagestyle{empty}

\begin{center}  
{\huge\bf\color{rossos} The collider landscape:
which collider for
establishing the SM instability?  }
{\huge\bf\color{rossos}}

\vspace{1cm}
{\bf\large Roberto Franceschini}$^a$, {\bf\large Alessandro Strumia}$^b$,  {\bf\large Andrea Wulzer}$^c$  \\[6mm]
{\it $^a$ Dipartimento di Matematica e Fisica, Universit\`a degli Studi di Roma Tre, Roma, Italia}\\[1mm]
{\it $^b$ Dipartimento di Fisica, Universit\`a di Pisa, Italia}\\[1mm]
{\it $^c$ Dipartimento di Fisica e Astronomia, Universit\`a di Padova, Italia}\\[1mm]

\vspace{1cm}
{\large\bf Abstract}\begin{quote}\large
Capabilities of future colliders are usually discussed assuming specific hypothetical new physics.
We consider the opposite possibility: that no new physics is accessible, and
we want to learn if the unnatural Standard Model is part of a vast landscape.
We argue that a main step in this direction would be 
establishing the possible instability scale of the Higgs potential.
This primarily needs reducing the uncertainty on the strong coupling and on the top quark mass.
We show that the top quark mass can be measured well enough
via a $t \bar t$ threshold scan with low $10^{33}\,{\rm cm}^{-2}{\rm sec}^{-1}$ luminosity, that seems achievable at
a `small'  $e^+ e^-$ collider in the LEP tunnel, or at the first low-energy stage of a muon collider.
\end{quote}
\end{center}

\setcounter{page}{1}
\tableofcontents

\section{Introduction}

Exploring short-distance physics systematically and conclusively has always been the true motivation for building colliders. However it is a fact that recent past colliders have been instead justified by safe Standard Model (SM) targets (such as finding the $Z$, Higgs, or the top quark) and by often overemphasized Beyond the SM (BSM) targets.  Maybe this happened because pure exploration is not considered a valid motivation for the community at large.
In this case, no collider will be built in the medium-term future, nor any other large-scale enterprise will be initiated aimed at probing short-distance physics, because we exhausted the SM targets and because the most compelling BSM ideas have lost a good portion of their appeal and motivation. 

\smallskip

If instead the quest for exploration will prevail, we will soon have to decide which collider or colliders to build, based on an unavoidably subjective assessment of the exploration potential of the different projects. The role of BSM theory in this context is to bring objective elements for the assessment by identifying possible BSM realities and quantifying the perspectives for probing them. For instance one might find interesting to establish if, or not, the observed Dark Matter abundance is due to the thermal freeze-out of an electroweak charged particle, and rank future colliders by the amount of progress they can make in answering this question. 
The ranking is objective, but in some cases it strongly depends on the question that is being asked, which is subjective. 
For instance, Dark Matter might be an electroweak multiplet with no strong interactions,
or a strong multiplet with no electroweak interactions: very different colliders are needed in the two cases.

\medskip

The need of studying capabilities of future colliders  from multiple perspectives, in order to offer the most complete assessment possible of their exploration potential, is well-understood in the community. The focus so far has been on BSM models or scenarios that foresee new physics at accessible energies. 

In this paper we consider instead the opposite possibility: that the apparent unnaturalness in the weak scale and in the cosmological constant originates from anthropic selection in a landscape of many vacua, possibly $N \sim 10^{500}$, populated by cosmological inflation~\cite{Weinberg:1987dv,hep-ph/9707380,hep-th/0603249,hep-th/0302219,1906.00986}.
In this context no new physics could exist up to inaccessibly high energies, possibly up to the Planck scale.
The appeal of this plausible although vague interpretation 
resides in its radical conservatism (standard theory brought to its extreme consequences can explain why the vacuum energy and the weak
scale seem unnatural), as opposed to conservative radicalism 
(invent unusual new theories arranged such that we do not see them yet). 

Its limitation is the lack of predictions and of concrete implications. 
We might be able to tell whether the scenario is true only after getting access to the microscopic theory that generates the landscape of vacua, by means of experiments we have no idea how to build. 
The theory itself is not known, and definitely not unique. 
It seems realized in the string/$M$-theory  framework,
altought huge landscapes of vacua also arise in Quantum Field Theories
with some hundreds of heavy scalars~\cite{hep-th/0501082,1911.01441}.

No collider, nor any other type of experiment envisaged so far
can falsify anthropic selection and the landscape. 
Still we can ask which collider would be the most useful from this unusual point of view. For this assessment we clearly have to depart from the usual practice of studying collider perspectives to discover or bound specific hypothetical new particles or interactions, 
since no new physics is needed at accessible energies in the scenario we are considering. 
Rephrasing in modern form the words sometimes attributed to Kelvin ({\em `there is nothing new to be discovered in physics now... all that remains is more and more precise measurements'})~\cite{Kelvin}\footnote{Kelvin's statement illustrates that predicting the future is remarkably difficult.}: maybe all that remains, for the near future, could be measuring the SM more and more precisely.

At the same time the landscape scenario provides a specific direction where this endeavour can have fundamental significance: {\em measuring more precisely the free parameters of the SM} that seem  `fundamental'  because we do not understand their values. Before discussing what it means, let us tell what it does not mean. The following example of measurements, while being relevant from different perspectives, are not relevant from this perspective:
\begin{itemize}
\item Measuring the Higgs interactions to fermions adds nothing, 
as the fundamental parameters of the SM that control these interactions, 
the Higgs Yukawa couplings, are already more precisely measured from the fermion masses. The same applies to all the single Higgs couplings and to the trilinear coupling. 
\item Similar considerations can be made for the $g-2$ of the muon and other very precise measurements that do not help determining the SM fundamental parameters.\footnote{Needless to say,
constants such as $c,\hbar, k_B$ are now understood as arbitrary conversion units.} \end{itemize}

What becomes relevant according to this point of view is acquiring more information about the fundamental constants that act as `coordinates' for the SM, if it lives in a landscape of \mbox{$N\gg 1$} vacua. 
If a plausible theory will become available for the microscopic structure of the landscape of vacua, 
the augmented knowledge will help us locate the single vacua we happen to live in, and in principle to identify it uniquely by asymptotically accurate measurements. 
Very roughly, if $N\sim 10^{500}$ we need to measure fundamental constants up to acquiring 500 digits of information. 
In Appendix~\ref{H} we discuss how Conditional Entropy allows
to quantify the required and the present amount of information.
Depending on the structure of the landscape theory, specific strategic measurements add more information. 
Since this structure is presently unknown the discussion remains abstract: 
we do not presently know which measurements would have strategic significance.
We reasonably guess that this happens for those parameters that happen 
to be close to a qualitative transition or anthropic selection boundary


A clear case are the parameters that control the possible existence of an extra minimum of the Higgs potential, 
that appears at field values much above the physical SM minimum where $\langle H\rangle=v=174$~GeV. 
The existence or not of this second minimum is a qualitative feature of the SM extrapolated up to high energies, 
which provides structural information relevant for future attempts of 
deriving the SM as one vacuum of a deeper high-energy quantum-gravity theory.
For instance, a negative Higgs quartic at the Planck scale might rule out entire classes of vacua,
or conversely select them if a deeper vacuum is obliged to exist (no de Sitter conjecture, see e.g.\ section 5 of~\cite{1505.04825}). 

The current measurements of the relevant parameters (namely, $\alpha_3$, $M_h$ and $M_t$) favour the existence of a second SM minimum, but more precise determinations are needed for a conclusive assessment. Furthermore, if the second minimum exists, precise measurements will allow us to compute the height and the width of the potential barrier between the two minima, which in turn is relevant for the cosmological history of the universe, possibly associated with an anthropic selection boundary. In fact, it is non-trivial that the $\langle H\rangle=v$ minimum is selected during inflation, nor that it is preserved by reheating~\cite{1112.3022,1505.04825}. The conditions for this to happen depend also on cosmological parameters, such as the Hubble scale during inflation and the reheating temperature. 
When the latter parameters will be also determined, precise measurements of $\alpha_3$, $M_h$ and $M_t$ will allow to establish if the conditions are satisfied
(see e.g.\ fig.~14 of~\cite{1505.04825} and fig.~5 of \cite{1706.00792}).
A possible anthropic origin of their values would arise
if the value of the parameters are close to the anthropic boundary defined by these conditions.

\medskip

The rest of this paper is organized as follows. 
In section~\ref{asmH}
we review the vacuum stability issue and the perspectives for future progress by improving the determination of the relevant parameters. 
We will see that the High Luminosity stage of the LHC (HL-LHC) can reduce the uncertainty on $M_h$  down to a sufficient level, while a fully satisfactory determination of $\alpha_3$ would require a factor 10 improvement of the current lattice precision, or a factor of 3 improvement of the sensitivity of the FCC-ee future collider (when theoretical uncertainties are taken into account) that is the most effective one for this measurement. Finally the measurement of $M_t$ can only be improved by a future lepton collider operating close to the threshold for top quark pair production. 
The requirement specifications for such collider, in terms of the integrated luminosity and of the beam energy spread that is required for a sufficiently accurate $M_t$ measurement, is studied in section~\ref{Mt}. 
The measurement can be performed at proposed future colliders, such as FCC-ee~\cite{Maier:2019vll}, CLIC~\cite{1807.02441,1303.3758,2103.00522} and ILC~\cite{1310.0563}, with a much wider scope that the top mass determination. We argue that unconventional sufficient alternatives for this specific measurement could be an $e^+e^-$ collider in the LHC tunnel (also known as LEP3) or a very compact first stage of a $\mu^+\mu^-$ collider. Finally, we report our conclusions in section~\ref{concl}.

\section{The SM vacuum (in)stability}\label{asmH}
As discussed in the introduction, measurements aimed at establishing the existence of a second minimum in the SM Higgs potential, such that the physical minimum is unstable, are of strategic relevance from the landscape perspective. In this section we summarize the current status and the future perspectives for the assessment of the (in)stability of the SM vacuum.

\subsubsection*{ The Higgs quartic coupling $\lambda$ extrapolated around the Planck scale}
\begin{figure}[t]
$$\includegraphics[width=\textwidth]{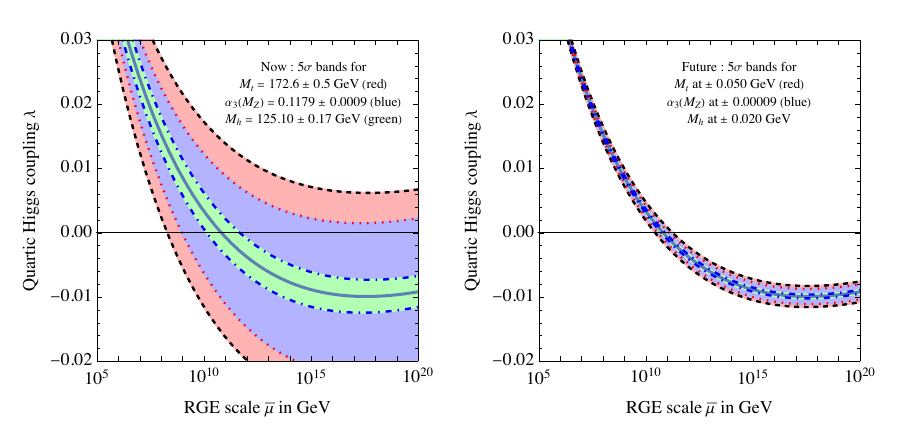}$$
\caption{\label{fig:run} Running of the quartic Higgs coupling, that determines the Higgs potential and
its instability scale. The $\pm 5\sigma$ bands associated with the uncertainty in the top quark mass $M_t$
are plotted as dashed gray, those associated to $\alpha_3(M_Z)$ as dotted red, 
those associated to $M_h$ as dot-dashed blue.
}
\end{figure}

The sign of the Higgs quartic coupling $\lambda$ extrapolated at the Planck scale $\mpl$ controls the existence of the second minimum. Its current central value is slightly negative, suggesting that the second minimum exists. Using the $\overline{\rm MS}$ scheme for $\lambda$ and $\alpha_3$  and the pole mass definition for $M_{t}$ and $M_h$, for values of the parameters close to the current central values, it equals
(see also~\cite{1307.3536})
\beq
\label{eq:lammp} 
\resizebox{0.99\textwidth}{!}{$\displaystyle
\lambda (\mpl ) \simeq  -0.0096- 0.0065\Mtdiff +0.0024\asdiff +0.0030\Mhdiff .$}
\eeq
The RG running of the quartic coupling is shown in fig.\fig{run} {(left panel)} including variations of 
$M_t$, $\alpha_3$ and $M_h$ within their present $\pm5\sigma$ ranges. The details of the evolution shown in fig.\fig{run} depend on the assumed central value. For $M_{t}$ we employ a central value close to the average of the LHC results, with an uncertainty corresponding to the latest published results (coming from the 7 TeV and 8 TeV runs) 
\beq M_t = \left\{ \begin{array}{ll}
172.69\pm 0.25_{\rm stat} \pm 0.41_{\rm syst} & \hbox{ATLAS, 20.2/fb~\cite{1810.01772}} \\
172.44\pm 0.13_{\rm stat} \pm 0.47_{\rm syst} & \hbox{CMS, 29.7/fb~\cite{1509.04044}}\\
174.30\pm 0.35_{\rm stat} \pm 0.54_{\rm syst}& \hbox{CDF, D0, $\le$9.7/fb~\cite{1608.01881}}\\
\end{array}\right.\eeq
For $\alpha_{3}$ and $M_{h}$ we pick the current PDG~\cite{Zyla:2020zbs} value and their uncertainties. 

With the current central values, the Higgs quartic is negative at high energy, but the opposite sign cannot be excluded at the $5\sigma$ level. Establishing firmly that the high energy Higgs quartic is negative is thus a clear first target for future investigations in this area. We see in the left panel of fig.\fig{run} that the current uncertainty on $M_h$ is sufficient for this task, and only a mild improvement on $\alpha_3$ is needed. The uncertainty on $M_t$ should instead significantly decrease, reaching $\delta M_t\approx 250~\MeV$. It should be emphasized that these estimates are strongly sensitive to the central value assumed by the parameters. For instance if $M_t$ was lower by 1~GeV, half of the measurement precision estimates above would not yet quite be sufficient to establish instability.

\smallskip

A second target is the determination of the scale at which $\lambda$ crosses zero and runs negative. This scale is largely uncertain, ranging from $10^{8}\GeV$ up to the Planck scale with the current $1\sigma$ errors of the SM parameters. It should be noted that the scale of crossing is gauge-dependent and thus unphysical. A physical gauge-independent definition of the scale of instability is provided instead~\cite{1307.3536,1505.04825} by the maximal height of the Higgs potential barrier, namely $\Lambda^4 = \max _h V_{\rm eff}(h)$.
This is given by~\cite{1307.3536}
\beq
\resizebox{0.94\textwidth}{!}{$\displaystyle
\log_{10}\frac{\Lambda}{\GeV} = 10.5 -1.3\Mtdiff + 0.6 \asdiff + 1.1\Mhdiff$}\,,
\label{eq:lambdai}
\eeq
where $M_{t}$ and $M_h$ are the pole top and Higgs masses and the $\overline{\rm MS}$ scheme is used for $\alpha_3$.

\begin{figure}[t]
$$\includegraphics[width=0.7 \textwidth]{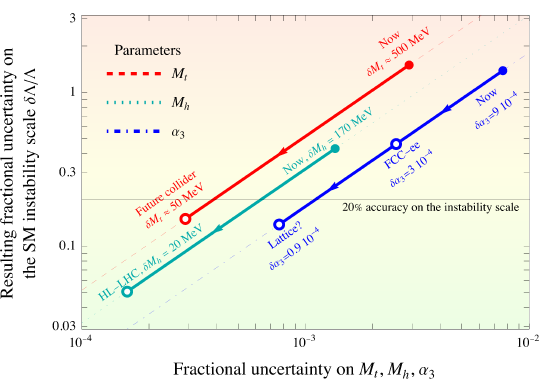}$$
\caption{\label{fig:target} 
The $1\sigma$ relative uncertainty on the scale of instability determined by eq.~(\ref{eq:lambdai})  as function of the relative precision of the measurements of $\alpha_{3}$, $M_{t}$, and $M_{h}$. The horizontal shade at 0.2 corresponds to a determination of  the instability scale at 20\% precision. The current situation and future improvements are marked as full and empty dots, respectively.}
\end{figure}

In order to assess the necessary precision to `measure' the instability scale we propagate uncertainties in eq.~(\ref{eq:lambdai}) and show in fig.~\ref{fig:target} the resulting relative uncertainty $\sfrac{\delta  \Lambda}{\Lambda}$ that corresponds to a given precision in the measurement of each of the three parameters $M_{t}$, $M_{h}$, and $\alpha_{3}$. The top mass and $\alpha_{3}$ are currently the largest sources of uncertainties, while $M_{h}$ is almost precise enough to determine the instability scale. Setting an arbitrary threshold of around $20\%$ precision on $\Lambda$, fig.~\ref{fig:target} shows that an absolute error of around $10^{-4}$ would be needed on $\alpha_3$, while the top mass should be known with error $\delta M_t=50\,\MeV$.

\subsubsection*{Improved $M_{t}$ determination prospects}
As the uncertainty on $M_{t}$ is reflected on the largest uncertainty on eq.~(\ref{eq:lambdai}), we start from discussing the prospects for progress in its determination. As we assume the validity of the SM up to very short length scales, the evaluation  of the performance of each experiments is evaluated under this assumption. 
Top quark loops affect various lower-energy observables, that are thereby sensitive to the top quark mass.
Previous work found that SM fits will not allow precise enough measurements of 
the top mass $M_t$~\cite{1508.05332}.
As a consequence, the only option to measure $M_{t}$ better is to measure it at future colliders. 

In this context the HL-LHC stands in a delicate position as the top quark sample is already large enough that systematic uncertainties dominates in analysis of the 7+8 TeV LHC data. 
Indeed, measuring the top quark mass summing the energies of its visible decay products 
is like measuring the pig mass summing sausages: higher statistics allows a better Monte Carlo modeling, but leaving systematics uncertainties untouched.
The present uncertainty about 500~MeV is at the limit to which  tools such as leading-log Monte Carlo parton shower generators are considered trustable. The inclusion of higher   perturbative orders  in the matrix elements attached to the present parton showers can improve this situation, but a measurement of $M_{t}$ with uncertainty comparable to $\Lambda_{\rm QCD}$ remains challenging. Thus we consider unlikely that the HL-LHC will improve the present uncertainty on $M_{t}$ by the substantial factor that is needed to firmly establish the scale of the SM instability.

\smallskip

Also enlarging the scope of HL-LHC to  `alternative' strategies for the top quark mass measurements, e.g. reviewed in~\cite{Corcella:2017rpt}, we find a limited improvement compared with the target imposed by our question. Even barring experimental uncertainties, the `alternative' methods are hitting the limitations of the present computations in describing effects commensurate with $\Lambda_{\rm QCD}$ either because of matching of fixed order and parton shower computations in the `alternative' observables\cite{FerrarioRavasio:2018ubr}, or uncertainties in the knowledge of hadronization physics~\cite{Corcella:2017rpt}, or possible lack of understanding of the effects of the colored environment~\cite{Argyropoulos:2014zoa} in which the short-distance $t\bar{t}$ are produced at LHC and so on. 
All in all, we will need a future collider beyond the HL-LHC to measure the top quark mass with sufficient precision to `measure' the instability scale. Great prospects are offered by $e^{+}e^{-}$ colliders, and in principle $\mu^{+}\mu^{-}$ colliders, that can determine the top quark mass from a center-of-mass energy scan around the threshold of the $\ell^{+}\ell^{-}\to t\bar{t}$ reaction around $2M_{t}$~\cite{1811.03950,1912.01275,2103.00522,1908.11299}. This will be discussed in section~\ref{Mt}.

\subsubsection*{Improved $\alpha_{3}$ determination prospects}
The present knowledge  of $\alpha_{3}$ results in a subdominant uncertainty on the instability scale compared to $M_{t}$. 
Still it is too large to draw conclusions on the fate of the quartic coupling at high energy. It can in principle be improved pursuing any of the presently employed techniques reviewed in~\cite{Zyla:2020zbs}. 

The most precise present determinations of $\alpha_3(M_Z)$ from experiments, with an uncertainty of order $\pm 0.0010$, come from low-energy $\tau$-decay and parton distribution functions fits. The determination from LEP has a larger uncertainty $\pm 0.0030$  
dominated by the uncertainty on the lepton/hadron ratio measured at the $Z$ pole,
$R_\ell=\Gamma(Z\to \hbox{hadrons})/\Gamma(Z\to\mu^+\mu^-)$,
affected at loop order by $\alpha_3$ and measured with $0.12\%$ uncertainty 
(mildly dominated by statistics).

According to FCC-ee studies~\cite{1308.6176,1512.05194,1806.06156,FCC:2018byv}, 
exploiting  $10^5$ times more $Z$ bosons than the full LEP data sample, 
the uncertainty on $R_\ell$ can be reduced by a factor 20 down to $\pm 0.005\%$,
so that the uncertainty on $\alpha_3$ can too be reduced by a factor 20, down to
$\pm 0.00016$, {barring any theory uncertainty~\cite{1512.05194}}. 
However by applying projected theory uncertainties onto this extractionof $\alpha_{3}$, Ref.~\cite{1512.05194} gives an estimated $\pm 0.00030$ for FCC-ee and $\pm 0.00070$ for ILC.
As the ILC is expected to yield a comparable number of $Z$ boson to `site filler' $e^{+}e^{-}$ machines (e.g. at FNAL~\cite{SiteFillerFNAL,SiteFillerEE}), we can take the ILC result as a ballpark indication for the possible results of such `site filler' project, or of LEP3~\cite{1112.2518,1208.0504} is ran at the $Z$-pole.

All in all, the high-energy $e^{+}e^{-}$  colliders in the most optimistic scenario can improve a factor around 20
compared to the present determinations from the same type of measurement, or,  expressed differently, a factor around 5 compared to the present world-average. From the current central value, the FCC-ee determination of $\alpha_{3}$ would be sufficient to establish the change of sign of the Higgs quartic at more than $5\sigma$, while  ILC or LEP3 would do that at around 5$\sigma$.  None of these measurements, however, would precisely pinpoint the scale of the instability, as shown in fig.~\ref{fig:target}. 

\smallskip

The present most precise single `measurement' of $\alpha_{3}$ is obtained from lattice QCD calculations of suitable quantities~\cite{PDGalpha3}. 
Based on the extrapolations of Ref.~\cite{Lepage:2014fla} on the reduction of the computing cost, and on their estimates of the impact of adding more perturbative orders, reduction of lattice spacing, and accumulated statistics, an improvement of the current lattice QCD uncertainty on $\alpha_{3}$ by a factor up to 10 can be expected in the next decade at fixed computing cost by adding one order in perturbation theory inputs in the lattice extraction of $\alpha_{3}$. More recent estimates on the progress of the lattice QCD determination of $\alpha_{3}$ contained in~\cite{2203.08271} give less optimistic prospects. 
Such factor 10 improvement would be sufficient, according to fig.~\ref{fig:target}, for a $20\%$ precision on the instability scale determination.

\subsubsection*{Improved $M_{h}$ determination prospects}

The Higgs boson mass as currently measured at the LHC has a small impact on the instability scale thanks to an already quite precise measurement. Knowledge of $M_{h}$ can be improved by future HL-LHC measurements in clean channels such as $h\to 4\ell$ and $h\to\gamma\gamma$. 
Exploiting a dataset larger by about two orders of magnitude, the HL-LHC could reach around 20~MeV uncertainty~\cite{Cepeda:2019klc}. This is well below the target for $20\%$ measurement of the scale $\Lambda$ (see fig.~\ref{fig:target}).

\section{A top threshold collider}\label{Mt}

\begin{figure}[t]
$$\includegraphics[width=0.4\textwidth]{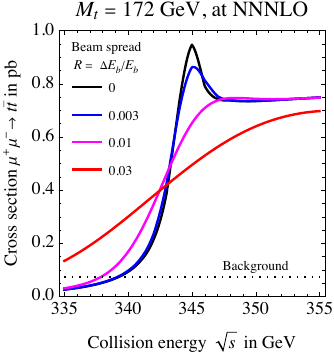}\qquad\qquad
\includegraphics[width=0.4\textwidth]{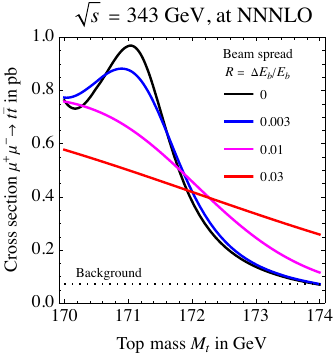}$$
\caption{\label{fig:Mtexp} 
Cross section for $\ell^+  \ell^- \to t\bar t$ around threshold at NNNLO accuracy and including Higgs-EW corrections, obtained with the {\tt{QQbar\_threshold}} code~\cite{1605.03010,1711.10429}. The background estimated for the CLIC analysis in~\cite{2103.00522}, of $73\,{\rm fb}$, is reported as a dotted line. {\bfseries Left}: as function of $\sqrt{s}$, fixed $M_t$. {\bfseries Right}: as function of $M_t$, fixed $\sqrt{s}$. We here employ the `potential-subtracted' top mass~\cite{Beneke:1998rk}.}
\end{figure}

The cross section  for $\ell^+\ell^-\to t\bar t$ around the kinematical threshold $\sqrt{s}\sim 2 M_t$ is strongly sensitive to the top mass~\cite{Fadin:1987wz,Fadin:1988fn,Strassler:1990nw,hep-ph/9801397,hep-ph/9802379,hep-ph/0107144,1712.02220}. The large variation of $\sigma$ with $\sqrt{s}$ is shown on the left panel of fig.\fig{Mtexp} (black curve), based on the NNNLO SM predictions from~\cite{1605.03010,1711.10429}. The spike is due to the $t\bar{t}$ $1s$ bound state. Correspondingly, as shown on the right panel of the figure, the cross section at fixed $\sqrt{s}$ depends strongly on $M_t$.\footnote{In the plot, and in the rest of this section, we employ the `potential-subtracted' top mass~\cite{Beneke:1998rk}.} Thanks to the high sensitivity to $M_t$
\beq \label{eq:gain}
\frac{d\ln \sigma}{d\ln M_t}\sim 1.6\, \frac{M_t}{\Gamma_t} \approx 200\,,
\eeq
a cross section measurement with modest $10\%$ relative accuracy enables a determination of the top quark mass at the $0.05\%$ level. The cross section close to the threshold is around $500$~fb, therefore a collider integrated luminosity as small as $\mathcal{L} \simeq0.2\, {\rm{fb}}^{-1}$ is sufficient for a $10\%$ statistical uncertainty on the cross section measurement.  In turn, based on the rough estimate above, this measurement could enable a determination of the top mass with $\delta M_t=5\,10^{-4} M_t=86$~MeV, close to the instability scale measurement target of $\delta M_t=50$~MeV, fig.~\ref{fig:target}. 

The vast literature (see e.g.~\cite{1603.04764,1604.08122,1807.02441,1902.07246,2103.00522,FCC:2018evy}) on $t\bar t$ threshold cross section measurements at lepton colliders considers a relatively large integrated luminosity (typically, $\mathcal{L} \sim100\,{\rm{fb}}^{-1}$), which can be available at colliders like CLIC, ILC, CEPC and FCC-ee, and quantifies the expected error on $M_t$ based on the properties (eminently, the shape in energy of the luminosity spectrum) of the specific collider project under examination. Here instead we want to assess the characteristics that a generic collider should possess, in terms of integrated luminosity and luminosity spectrum, for a sufficiently accurate determination of $M_t$. 

\smallskip

Furthermore, existing studies of the top threshold have a broader target than the determination of $M_t$, including independent measurements of the top quark width $\Gamma_t$ and of the top Yukawa coupling $y_t$. 
From our perspective instead, $\Gamma_t$ and $y_t$ are predicted by $M_t$ and the other SM parameters, because no new physics exists to modify the SM relations. 
All these parameters can be independently accurately determined with $\mathcal{L} =100\,{\rm{fb}}^{-1}$ and a scan on the center of mass energy $E_{\rm{cm}}$ of the collider at ten points with equal integrated luminosity, spaced by 1~GeV, which is the baseline running scenario for most of these studies. This scan setup is not far from optimal~\cite{2103.00522} for the simultaneous determination of all these parameters, but expectedly not so (see~\cite{2103.00522} for a discussion) if the only target is the $M_t$ determination as in our case. A reassessment of the scan strategy is needed.

Apart from the luminosity, the most important feature that a top threshold collider must possess for an accurate $M_t$ determination is a luminosity spectrum that is narrowly localized around the nominal collider energy, $\sqrt{s}\simeq E_{\rm{cm}}$. We model the spectrum with a Gaussian centered at $E_{\rm{cm}}$ and standard deviation $R/\sqrt{2}\,E_{\rm{cm}}$, with $R=\Delta E_b/E_b$ the energy spread of each beam. 
The relative standard deviation $R/\sqrt{2}$ should be compared with the width of the cross section shape on the left panel of fig.~\fig{Mtexp}, which in turn can be estimated as $\Gamma_t/M_t$. If it is much smaller than that, namely if $R\ll\Gamma_t/M_t\sim1\%$, the cross section is not affected by the convolution and the shape of $\sigma(E_{\rm{cm}},M_t)$ displays the strong dependence on $E_{\rm{cm}}$ and $M_t$ previously described. On the other hand, if $R$ is larger the dependence is smoothed out significantly, as shown by the colored lines in fig.~\ref{fig:Mtexp}, entailing a reduction of the $M_t$ measurement precision.

The heuristic considerations above can be summarized in the following estimate for the error
\beq\label{eq:dMtexpected}
\delta M_t \approx \frac{\Gamma_t}{1.6\sqrt{N_t}}
\left[1+\left( \frac{M_t R}{0.5\,\Gamma_t} \right)^2\right]^{p} ,\qquad
N_t = \mathcal{L}  \med{\sigma},\qquad\med{\sigma}\approx 0.5\,{\rm pb}\,,
\eeq
where $N_t$ is the total number of produced $t\bar t$ events. The scaling with $1/\sqrt{N_t}$ is dictated by the statistical accuracy in the cross section measurement, and the prefactor is chosen according to eq.\eq{gain}. The factor 0.5 arises matching a Breit-Wigner with a Gaussian. The power $p \approx 0.45$ 
arises in view of the specific shape of $\sigma(E_{\rm{cm}},M_t)$, and mostly depends on how the sensitivity in eq.\eq{gain} is reduced at large $R$.

\smallskip

\begin{figure}[t]
$$\includegraphics[width=0.45\textwidth]{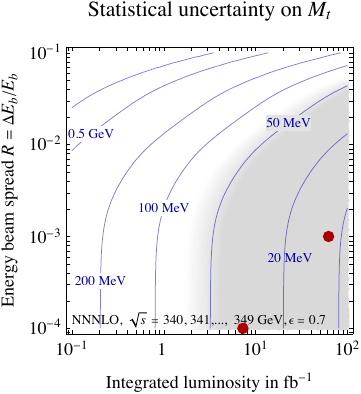}\qquad\includegraphics[width=0.45\textwidth]{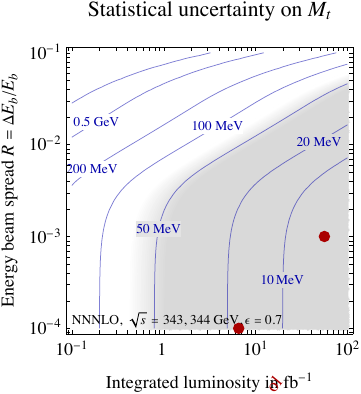}$$
\caption{\label{fig:Mtexp2} 
{\bfseries Statistical uncertainty on the top mass}.
Initial State Radiation is neglected, as appropriate for a muon collider.
The left panel assumes running at 10 values of
$E_{\rm{cm}}=\{340,341,\ldots,349\}\GeV$ with $\mathcal{L} /10$ luminosity at each point.
The right panel assumes running at  $E_{\rm{cm}}=\{342,343\}\GeV$ with $\mathcal{L} /2$ luminosity at each point.
The results are reported in the plane formed by the beam energy spread $R$, and the luminosity $\mathcal{L} $.
We assumed a $70\%$ efficiency for $t\bar t $ reconstruction. 
In the shaded region the systematic uncertainty on $M_t$ estimated in eq.\eq{Mtsyst} is larger than the statistical uncertainty.
}
\end{figure}

For a solid estimate of the uncertainty we include the smearing due to the energy spread in the cross section predictions and we performe a $\chi^2$ fit to the top mass using measurements at several $E_{\rm{cm}}$ points. A constant background of  $73$~fb, from the CLIC analysis in~\cite{2103.00522}, is considered in the analysis. The results are displayed in fig.~\ref{fig:Mtexp2} as $\delta M_t$ contours in the $(\mathcal{L} ,R)$ plane.  The figure assumes $70\%$ efficiency in the reconstruction of the two top quarks~\cite{2103.00522}.
The effects of Initial State Radiation (ISR) of photons, which are significant for $e^+e^-$ colliders as we will see, are not 
included in the figure. Only statistical uncertainties are included in the fit. We  later comment on the expected systematic and parametric uncertainties.

\smallskip

The left panel of fig.~\ref{fig:Mtexp2} shows the results for ten collider $E_{\rm{cm}}$ points, equally spaced by 1~GeV starting at 340~GeV, with equal luminosity, of $\mathcal{L} /10$, collected at each run. The precision on $M_t$ is significantly inferior than the one estimated by eq.\eq{dMtexpected}, and correspondingly a higher luminosity is needed to attain a given $\delta M_t$ target. For instance $\delta M_t=50$~MeV, for negligible $R$, requires more than $3\,{\rm{fb}}^{-1}$, while $0.8\,{\rm{fb}}^{-1}$ would suffice according to eq.\eq{dMtexpected} including the $70\%$ efficiency. This is because the threshold scan points are not optimized for the sensitivity to $M_t$, as previously explained. The best results would be obtained by collecting the entire luminosity at the single point that maximizes the sensitivity. For a true value of $M_t=172$~GeV, which we assume for our analysis, the optimal point would be at $E_{\rm{cm}}=343.5$~GeV, nearly independently of the beam energy spread. However with a single energy point the $\chi^2$ often displays a secondary minimum, and furthermore a running scenarios with multiple energy points is arguably favored for the reduction of systematic uncertainties that are correlated at the different points. 

We thus consider two energy points spaced by 1~GeV, whose optimal positions are found to be at 333 and at 334~GeV. This configuration improves the result significantly, as shown on the right panel of fig.~\ref{fig:Mtexp2}. The improvement is less pronounced at large $R$, because the beam energy spread flattens out the dependence of the cross-section on $E_{\rm{cm}}$, asymptotically making all the points in the threshold region equally sensitive to $M_t$. The right panel of the figure is in good agreement with the estimate in eq.\eq{dMtexpected}.

\begin{figure}[t]
$$\includegraphics[width=0.4\textwidth]{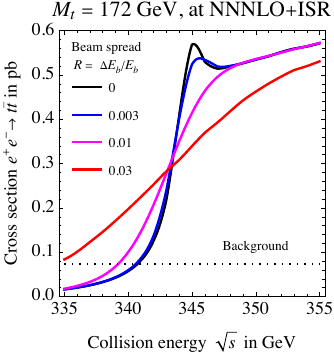}\qquad\qquad
\includegraphics[width=0.4\textwidth]{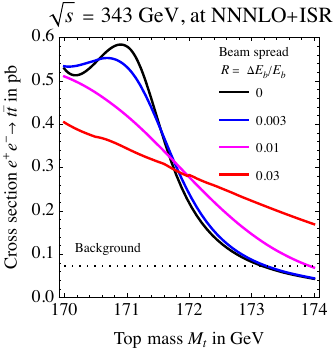}$$
\caption{\label{fig:MtexpISR}
As in fig.~\ref{fig:Mtexp}, but with Initial State Radiation included, as appropriate for a $e^+ e^-$ collider.}
\end{figure}

\smallskip

The scan  optimization depends on the true value of $M_t$, especially when $R$ is small, 
since the optimization is less relevant for large beam energy spread as previously explained. The true top mass is uncertain. 
Therefore the luminosity estimates on the right panel of the figure should be taken with care, while the ones on the left panel are more robust to the true value of $M_t$. On the other hand the true value of $M_t$ can be determined with increasing accuracy during the collider operation, enabling an improved determination of the optimal points. Some tests we performed assuming the top mass in a safe confidence range, chosen on the basis of the present-day error, suggest that around twice the luminosity estimated with two optimized scanning points are sufficient to match the accuracy. For instance, $2\,{\rm{fb}}^{-1}$ is enough for $\delta M_t=50\,$MeV at small $R$. For larger $R\gtrsim10^{-2}$, $\delta M_t=50\,$MeV requires more luminosity, but still below $10\,{\rm{fb}}^{-1}$.

\subsubsection*{Initial state radiation}

\begin{figure}[t]
$$\includegraphics[width=0.45\textwidth]{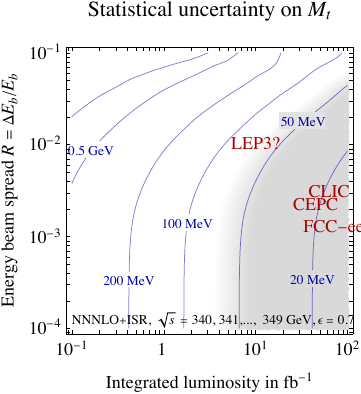}\qquad\includegraphics[width=0.45\textwidth]{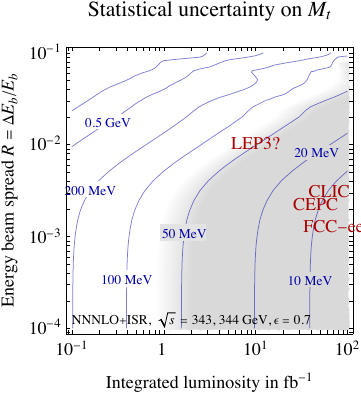}$$
\caption{\label{fig:Mtexp2ISR} 
As in fig.~\ref{fig:Mtexp2ISR}, with Initial State Radiation effects included,
as appropriate for a $e^-e^+$ collider.
}
\end{figure}

The estimates presented so far apply to a muon collider, while for $e^+e^-$ colliders the low mass of the electron entails a significant impact of photons ISR on the cross section, shown in fig.~\ref{fig:MtexpISR}. 
The effect of ISR is doubly negative. It lowers the value of the cross section close to the threshold and it reduced its sensitivity to the top mass. The result, 
shown in  fig.~\ref{fig:Mtexp2ISR}, is an increase of the required luminosity of a factor almost 3 at small $R$. The increase is less prominent when the beam energy spread is larger, so that ISR has a relatively milder impact on the collision energy spectrum.

Our results for 10 energy points, including ISR, are in good agreement with the literature. For instance, the FCC-ee collider with $\mathcal{L}=200\,{\rm{fb}}^{-1}$ and 
beam energy spread $R = 2~10^{-3}$~\cite{Zyla:2020zbs} could measure $M_t$ (when $\Gamma_t$ is varied with $M_t$ according to the SM relation) with $\pm 9 \MeV$ statistical uncertainty~\cite{FCC:2018evy}. CLIC could reach $\pm21\MeV$~\cite{1807.02441} with the half luminosity and energy spread $R=2~10^{-3}$~\cite{CLICspread}. The CLIC performances are slightly inferior, and in less good agreement with our estimate, possibly because of the beam-beam interaction effects, typical of linear colliders, that slightly reduce the luminosity in the threshold region.

\subsubsection*{Systematic and parametric uncertainties}

Several sources of uncertainty should be included for a realistic analysis, in addition to the statistical uncertainty we have estimated. First, one should consider experimental systematic uncertainties in the measurement of the cross section due to the uncertainties on the determination of the luminosity and of the $t\bar t$ acceptance and reconstruction efficiency, or on the background estimate. These uncertainties should be compared with the statistical uncertainty on the cross section measurements, which is modest (above one percent) for $\mathcal{L} \lesssim10\,{\rm{fb}}^{-1}$. Experimental systematics are thus expected to play a minor role.

\smallskip

Uncertainties also emerge from the imperfect knowledge of the beam energy distribution, namely of the central value $E_{\rm{cm}}$ and of the relative spread $R$ of the distribution. The central value is the reference scale for the measurement, therefore it should be known better than $\delta M_t/M_t$ ($0.03\%$, for $\delta M_t=50\,$MeV) not to impact our findings. The uncertainty on $R$ has no effect when $R$ is negligible but it can impact our results significantly when $R$ is of order few per mille or larger and the cross section is affected at order one by the convolution with the beam spectrum. An $R$ determination as accurate as $\delta M_t/M_t$ is expectedly needed in the large $R$ regime. This aspect was investigated in details for linear colliders in~\cite{1309.0372} (see also~\cite{1303.3758}), showing that the beam energy distribution parameters can be measured in Bhabha events with enough precision not to not affect the top mass determination. The conclusion might hardly be different for circular $e^+e^-$ colliders, but the point should probably be reassessed in the case of muon colliders.

\smallskip

The effect of systematic and parametric uncertainties on the cross section prediction is extensively studied in the literature~\cite{1603.04764,1807.02441,2103.00522,2203.06520}. They emerge from theory uncertainties on the resummation of threshold corrections, as well as from fixed order scale variations which alone amount to around 40~MeV~\cite{1603.04764}. Current uncertainties from $\alpha_3$ are expect to become negligible with the improvement on the $\alpha_3$ determination envisaged in the previous section. 
We thus expect systematic uncertainty on $M_t$ of around 
\beq\label{eq:Mtsyst} 
\delta M_t|_{\rm syst}  \approx  40 \MeV.
\eeq

A total uncertainty $\delta M_t=50\,$MeV, that is expected to settle the vacuum stability question conclusively as discussed in Section~\ref{asmH}, should thus be feasible. The results of the present section show that a top threshold collider with moderate beam energy spread could attain comparable statistical precision with few/fb integrated luminosity. This is a factor of 20 less luminosity than the one foreseen at the currently proposed top threshold colliders.\footnote{These colliders have broader scope than the measurement of $M_t$, which justifies the higher target luminosity.}

\subsubsection*{Unconventional top threshold colliders}

The ILC, CLIC and FCC-ee are mature well-studied collider projects that can match (and overcome) the $\delta M_t=50\,$MeV statistical precision target as previously discussed. We devote the rest of this section to speculate on the feasibility of the top mass measurement at unconventional colliders that might (or not) be convenient to build given financial or strategic considerations. Specifically, we consider an extension of the `LEP3' $e^- e^+$ proposed future collider, and a muon collider.

\begin{table}[t]
$$
\begin{array}{cl|cccccc}
\multicolumn{2}{c|}{\hbox{Collider}}& \hbox{LEP} & \hbox{LEP3} & \hbox{FCC-ee~\cite{FCC:2018byv}} & \hbox{CEPC~\cite{CEPC} } \\ 
\multicolumn{2}{c|}{\hbox{Total length $L$}} &26.6\km &26.6\km&100\km&100\km 
\\ 
\hline
 \rowcolor[rgb]{0.99,0.97,0.97}
Z & \hbox{$E_{\rm{cm}}=91\GeV$}  &   \sim 0.004&7^{*}&   460 & 115 \\  
 \rowcolor[rgb]{0.93,0.99,0.97}
W^+W^- & \hbox{$E_{\rm{cm}}=160\GeV$}  & \sim 0.01& 2^*&  56&  16 \\
 \rowcolor[rgb]{0.93,0.99,0.91}
Zh& \hbox{$E_{\rm{cm}}=240\GeV$}  & 0&1\hbox{~\cite{1208.0504}}&  17&  5 \\
  \rowcolor[rgb]{0.99,0.99,0.93}
t\bar t &  \hbox{$E_{\rm{cm}}=350\GeV$} & 
0 & 0.1^*&  3.8  & 0.5 \\  \rowcolor[rgb]{0.97,0.97,0.93}
\end{array}
$$
\caption{Expected luminosities in $10^{34}\,{\rm{cm}}^{-2}{\rm{s}}^{-1}$. 
The * stands for our estimate, based on rescalings from available literature.
\label{tab:coll}
}
\end{table}

LEP3~\cite{1112.2518,1208.0504} is a possible circular $e^- e^+$ circular collider in the existing LHC tunnel (previously used for LEP) with length $L = 26.6\,{\rm km}$. Multiple advances in accelerator physics 
enable 
to reach 240~GeV with $10^{34}{\rm{cm}}^{-2}{\rm{s}}^{-1}$ instantaneous luminosity, as in Table~\ref{tab:coll}, to measure the Higgs couplings precisely. The possibility of operating LEP3 at the top threshold $E_{\rm{cm}}\simeq350\,\GeV$ has not been studied, and is challenging for multiple reasons. Rough estimates of the conceivably achievable luminosity, merely based on the power emitted by synchrotron radiation, can be obtained as follows:
\begin{itemize}

\item { Rescaling LEP3 at $E_{\rm{cm}}=240\GeV$}. The power emitted by $N$ circulating $e^\pm$ in a ring or radius $r=L/2\pi$ is
\beq \label{eq:Wirr}
W_{\rm irr} = \sfrac{N e^2 (E_{\rm{cm}}/2)^4 }{3\pi  r^2 m_e^4 }\,.
\eeq
This should be smaller than about 100~MW, limiting to $\Lag_{Zh} \simeq 10^{34}{\rm{cm}}^{-2}{\rm{s}}^{-1}$~\cite{1112.2518,1208.0504,Zanetti} the LEP3 luminosity at $E_{\rm{cm}}=240\GeV$. The luminosity scales as $\Lag \propto E_{\rm{cm}} N^2$,  where the factor of $E_{\rm{cm}}$ accounts for the relativistic shrinking of bunches. Therefore with the same radiated power we estimate
\beq 
\Lag^{\rm LEP3}_{t\bar t} \approx \Lag^{\rm LEP3}_{Zh} (240/350)^7 \approx
0.88~10^{33}{\rm{cm}}^{-2}{\rm{s}}^{-1}\,,
\eeq
for the LEP3 collider at the top threshold $E_{\rm{cm}}=350\,\GeV$.

\item { Rescaling FCC-ee/CEPC at $E_{\rm{cm}}=240\GeV$}. The planned FCC-ee and CEPC circular collider with a length $L = 100$ km produce $\Lag_{\rm FCC-ee}\approx 4 \, 10^{34}{\rm{cm}}^{-2}{\rm{s}}^{-1}$~\cite{FCC:2018evy} and $\Lag_{\rm CEPC}\approx 0.5 \, 10^{34}{\rm{cm}}^{-2}{\rm{s}}^{-1}$~\cite{CEPC} instantaneous luminosity at $E_{\rm{cm}}=350\,\GeV$. The luminosity scales as $\Lag \propto N^2/L$ with the collider circumference, owing to the reduction of the collision frequency with $L$. 
Assuming again that the limiting factor is the total radiated power of eq.\eq{Wirr} we thus find that the luminosity achievable in the smaller existing LEP tunnel is
\beq 
\Lag_{t\bar t}^{\rm LEP3}\approx \Lag_{t\bar t}^{\hbox{\scriptsize FCC-ee}}/3.76^{3} \approx 0.75~10^{33}{\rm{cm}}^{-2}{\rm{s}}^{-1}\,,
\eeq
in good agreement with the previous estimate. Obviously a factor $8$ less luminosity would be obtained by rescaling CEPC.
\item A  similar estimate,  $\Lag_{t\bar t}^{\rm LEP3}\approx 2~10^{33}{\rm{cm}}^{-2}{\rm{s}}^{-1}$ is obtained rescaling in both radius and energy the claim $\Lag_{Zh}^{\rm 16~km}\approx 5.2~10^{33}{\rm{cm}}^{-2}{\rm{s}}^{-1}$, for a 100 MW collider with $L=16\km$~\cite{Sen}.

\end{itemize}
As we rescaled collider claims optimized for different lengths and energies, possible adaptations of the collider parameters (number of bunches, $\beta$, emittance) could lead to a mildly higher luminosity. For example the luminosity scales with a milder $1/E^{1.8}$ up to when beamstrahlung effects become relevant~\cite{Wenninger,Zimmermann:2015gga}. 

A significant challenge for the LEP3 collider operating at the $t\bar t$ threshold is the large beam energy loss of 28 GeV per turn, $(175/104.4)^4 \approx 8$ times higher than at LEP with $E_{\rm{cm}}=104.4\GeV$. This needs to be compensated by accelerating cavities $dE/dx$ in a few $\%$ fraction of the ring circumference. LEP used energy gradient $dE/dx = 7 \, {\rm MeV/m}$, but $dE/dx=45$ MeV/m is now possible~\cite{1701.06077}. Even if the total energy loss in one turn could be compensated, it would still be challenging to maintain the beam in a stable orbit while the energy is emitted. If all these challenges could be successfully addressed, a luminosity of order $\Lag=10^{33}{\rm{cm}}^{-2}{\rm{s}}^{-1}$ could be realistic as previously estimated. This corresponds to an integrated luminosity $\mathcal{L} =10\,{\rm{fb}}^{-1}$ for one year run. 

\medskip

We are obviously not in the position to estimate the beam energy spread that could be possibly attained, that depends in a non-trivial way on the energy and on the machine optics, and on the deployment of appropriate monochromatization techniques that allow a trade between the energy spread and luminosity. Based on our results in fig.~\ref{fig:Mtexp2ISR}, a percent-level beam energy spread (ten times worse than LEP, where $R\approx10^{-3}$) would be needed to attain the $\delta M_t=50$~MeV target if $\mathcal{L} =10\,{\rm{fb}}^{-1}$. 

\medskip

We finally comment on the possibility of a $\mu^+\mu^-$ top threshold collider. This option could be considered as a possible first stage of a future very high energy muon collider of $E_{\rm{cm}}=10\,\TeV$ or more~\cite{1901.06150}, that is currently being investigated by the International Muon Collider Collaboration (IMCC)~\cite{IMCC}. Such `First Muon Collider' was actually proposed long ago~\cite{hep-ph/9712486}~(see also~\cite{Barger:1997yk}).~\footnote{The uncertainty estimated in~\cite{hep-ph/9712486} for $100\,{\rm{fb}}^{-1}$ is in good agreement with ours, taking into account that a $t{\bar{t}}$ efficiency $\epsilon=(0.3)^2$ (much lower than the realistic $\epsilon=0.7$~\cite{2103.00522} we employ) is assumed in~\cite{hep-ph/9712486}. Furhtermore, the NNNLO cross-sections we employ give better sensitivity than the ones at NLO used in~\cite{hep-ph/9712486}.} Two parameter sets are proposed in~\cite{Palmer}. The first one with energy spread $R=10^{-4}$ and $\Lag_{\rm MuC}= 7\times 10^{32}{\rm{cm}}^{-2}{\rm{s}}^{-1}$, the second with $R=10^{-3}$ and $\Lag_{\rm MuC}=6\times 10^{33}{\rm{cm}}^{-2}{\rm{s}}^{-1}$. The total length $L$ of this collider would be $L=700\,{\rm{m}}$. Figure~\ref{fig:Mtexp2} shows that, in one year run, both options could achieve better precision than $\delta M_t=50$~MeV if systematic uncertainties could be reduced. Notice that the sensitivity of muon colliders (in fig.~\ref{fig:Mtexp2}) is slightly better than the one of $e^+e^-$ colliders with the same luminosity and energy spread because the of the absence of ISR.

\section{Conclusions}\label{concl}
The lack of new physics that keeps the weak scale and the vacuum energy naturally small
suggests anthropic selection in a landscape of vacua, and thereby
the possibility that no new physics exists within the reach of  next colliders.
We provided a first assessment of the potential of future colliders under this assumption.
In this situation, all what experiments can (and must) do is to measure the fundamental input parameters of the SM with increasing accuracy. 
If the SM is part of a landscape of vacua, 
accurate measurements will generically help to test
if it is part of a landscape, and locate it. {{While the concrete deployment of this program requires information on the detailed structure of the landscape theory that we do not currently possess, we argued}} that strategic measurements are those needed to assess the existence of a second minimum in the Higgs potential for Planck-scale values of the Higgs vacuum expectation value, namely the determinations of $M_t$, $\alpha_3$ and $M_h$. 

In Section~\ref{asmH} we defined accuracy targets for the measurement of these quantities, based on two distinct criteria. The first criterion, more loose,  is that we would like to establish with $5\sigma$ `certitude' that the second minimum exists, 
as suggested by the current central values. The second criterion, more ambitious, is the request to be able to perform a measurement of the scale of SM vacuum instability (defined, for example, as the maximal height of the potential barrier) 
with some reasonable accuracy, say $20\%$.

\smallskip

The most ambitious accuracy target for the measurement of $M_h$ can be attained at the HL-LHC. The target accuracy for $\alpha_3$ is a factor of 3 lower than what the most precise future collider project (FCC-ee) could achieve, and it is a factor 10 lower than the current lattice QCD determination. 
Such improvement from the lattice has been claimed to be possible in the literature, but also the opposite has been claimed. 

The situation for $M_t$ is more clear. The target uncertainty $\delta M_t=50$~MeV can be definitely (and only) obtained building a lepton collider operating at the top threshold. Several well-established future colliders such as FCC-ee, CLIC, CEPC and ILC can attain this target. They can actually achieve a smaller statistical error on $M_t$, which however hits systematic uncertainties that can be hardly reduced below 40~MeV. This singles out $\delta M_t=50$~MeV as the target for top mass determination regardless of vacuum instability considerations. 

In Section~\ref{Mt} we revisited the top mass determination from $\ell^+\ell^-\to t{\bar{t}}$ cross section measurement close to the top threshold, with the aim of identifying the minimal specification requirements for a lepton collider to measure $M_t$ with error $\delta M_t=50$~MeV. We pointed out that the integrated luminosity ${\mathcal{L}}\approx100\,{\rm{fb}}^{-1}$ that is generically assumed for the threshold scan is unnecessary for $\delta M_t=50$~MeV precision. A luminosity of few ${\rm{fb}}^{-1}$ is sufficient if the purpose, like in our case, is to measure only the fundamental SM parameter (i.e., the top Yukawa coupling) associated with $M_t$. A larger luminosity might be needed for an independent determination of other parameters like $\Gamma_t$, which is relevant only as a probe of new physics, which however we assumed not to exist. 

Our results might offer a guidance to the design of `unconventional' top threshold colliders to be built in place or in addition to the top-threshold stage of the other projects, if strategically convenient. One parameter that controls the quality of the beam for the measurement of $M_t$ is the beam energy spread $R$, whose optimal value is around $10^{-3}$ or below. The other parameter is the nature of the beam. 
In particular muon beams are favored over electrons because Infrared State Radiation effects reduce the sensitivity. We briefly discussed the possibility of upgrading the LEP3 $e^+e^-$ collider up to the top threshold, and of building a compact $\mu^+\mu^-$ dedicated collider. This latter option could be considered as a possible `demonstrator' stage of a future very high energy muon collider. Dedicated studies are needed.

\appendix\small

\section{Increasing the information content of known physics}\label{H}

More precise measurements of the SM parameters allow us to locate more precisely the SM in the landscape of vacua, i.e.\ to gain information on the specific vacuum `$\va$' realized in our universe. 
Notions from information theory can thus be employed to quantify the information gain. An attempt in this direction is presented in this appendix.

\subsection{General information-theory discussion}
In order to proceed, we label vacua by an integer $\va=\{1,\ldots,N\}$, 
and interpret it as instances of a statistical variable `$V$'. 
In the most uncertain situation, we could be in each vacuum $\va$ with probability equal probability $\wp(\va) = 1/N$. Therefore the Shannon entropy of the variable $V$ that describes the landscape is
\beq 
H({{V}}) = - \sum_{\va=1}^{N} \wp(\va) \ln \wp(\va) =\ln {N}\,.
\eeq
The Shannon entropy tells the number of information digits in the basis of the Euler number ($e$-digits) that must be measured in order to identify the vacuum uniquely among the $N$ options. 
For instance if $N=10^{500}\simeq e^{1200}$, the number of required $e$-digits equals $\ln{N}\simeq 1200$. When adding information by means of measurements, only the vacua similar to the SM remain compatible with available information and less digits of information will be needed.

We gain information by measuring the $n$ fundamental parameters
that characterize the effective QFT for the light particles of each vacuum:
the gauge, Yukawa and quartic couplings, and dimensional parameters.
We collectively denote them as $Y_i$ with $i=\{1,\ldots, n\}$. 
The value of the parameters is predicted to be $y_{\va,i}$ in each specific vacuum $\va$. 
Assuming (for simplicity) uncorrelated Gaussian distributed measurements with standard deviations $\sigma_i$, 
the probability of the measurements  giving as outcome the central values $y_i$ (namely to observe $Y_i=y_i\pm\sigma_i)$
assuming vacuum $v$ is
\bea
&&\wp(\vecf{y}\,|\va)=\prod_{i=1}^n \frac{1}{\sigma_i \sqrt{2\pi}} \exp\left[-\frac12 \frac{(y_{i}-y_{\va,i} )^2}{\sigma_i^2}\right]\, .\label{pyv}\eea
Using the (lack of) prior information about which vacuum is physical (i.e., $\wp(\va) = 1/N$)
\bea
&&\wp(\vecf{y})=\sum\limits_\va \wp(\vecf{y}\,|\va) \wp(\va)=\sum\limits_\va \frac1{N} \wp(\vecf{y}\,|\va)\,.\label{py}
\eea
After the measurements, the probability of each vacuum $v$ is  obtained by the Bayes theorem
\beq\label{postp}
\wp(\va|\vecf{y})=\frac{\wp(\vecf{y}\,|\va) \wp(\va)}{\wp(\vecf{y}\,)}=\frac{\wp(\vecf{y}\,|\va)}{\sum_\wa \wp(\vecf{y}\,|\wa)}\,.
\eeq
The Shannon entropy for the landscape variable $V$ after the measurements thus becomes
\beq\label{ShenM}
H({{V}}|\vecf{Y}=\vecf{y}) = - \sum_{\va=1}^{N} \wp(\va|\vecf{y}\,) \ln \wp(\va|\vecf{y}),
\eeq
smaller than the value it had before the measurements, $H(V)=\ln{N}$. 
Accurate enough measurements are needed in order for the entropy to decrease significantly. This can be seen by noticing that in the limit $\sigma_i\to\infty$ of very inaccurate measurements the posterior probability $\wp(\va|\vecf{y})$ in eq.~(\ref{postp}) approaches the flat prior probability $\wp(\va)=1/N$, because $\wp(\vecf{y}\,|\va)$ is independent of $\va$. More specifically, $\wp(\va|\vecf{y})$ can depart from the flat distribution significantly, such that the entropy decreases, 
only if $\sigma_i$ is smaller than the interval span by the parameter predictions $y_{\va,i}$ across the vacua. 
Namely, the measurement should be precise enough to discriminate among the vacua.

Eventually the distribution becomes localized on some vacuum $\bar\va$, $\wp(\bar\va|y)=1$, as $\sigma_i$ decreases further and becomes much smaller than the minimal separation between the $y_{\va,i}$ predictions at different vacua. Such precise measurements identify the vacuum uniquely reducing the entropy to zero, $H({{V}}|\vecf{Y}=\vecf{y})=0$, meaning that no additional information is required.  

\smallskip

The entropy in eq.~(\ref{ShenM}) depends on the values $y_i$ of the parameters that have actually been measured. Generically the entropy is smaller, for given measurement errors, if the measured values $y$ fall in a region that is less densely populated by the landscape. Indeed in that case the minimal separation between the different predictions is larger, and thus it is easier to identify the vacuum. Measurements of parameter values that are atypical in the landscape strategically provide more information. A well known limiting case, dubbed   `swampland', arises if the landscape leaves empty regions of parameter space. Observation of parameters in an empty region would
falsify the theory that generates the landscape.

\smallskip

However we do not know the parameter density in the vacuum, therefore we cannot exploit the knowledge of the measured values $y_i$. We can nevertheless proceed as follows. If the landscape theory is correct the measured parameter values will more likely be the `typical' ones predicted by the landscape statistics. 
Therefore it makes sense to average eq.~(\ref{ShenM}) over $y_i$, obtaining what is known as \emph{conditional entropy} in information theory
\beq
H({{V}}|\vecf{Y}\,) = -
\int d^ny 
 \sum_{\va=1}^{N} \wp(\vecf{y}\,) \wp(\va|\vecf{y}\,) \ln \wp(\va|\vecf{y})
 =-\int d^ny 
 \sum_{\va=1}^{N} 
 \frac1{N}
 \wp(\vecf{y}\,|\va) \ln \wp(\va|\vecf{y})
 \,.
\eeq
The expected average {\em information gained} by a generic measurement (without knowing its outcome) is then
\beq\label{eq:deltaH}
\Delta{H}=H({{V}}) -H({{V}}|\vecf{Y}) = \int d^ny 
 \sum_{\va=1}^{N} \frac1{N}\wp(\vecf{y}\,|\va) \ln  \frac{\wp(\vecf{y}\,|\va)}{\wp(\vecf{y})}
 \,.
\eeq

The dependence on the measurement uncertainties of the conditional entropy, and in turn of $\Delta{H}$, is qualitatively the same of the posterior Shannon entropy in eq.~(\ref{ShenM}). Namely, there is no information gain (i.e., $\Delta{H}=0$) if the measurement uncertainties $\sigma_i$ are much larger than the range of variability of the parameters in the landscape. The vacuum is fully determined (i.e., $\Delta{H}=H(V)=\ln{N}$) if $\sigma_i$ is much smaller than the typical spacing among the parameter predictions in the different vacua. The information gain increases as the measurement accuracy improves, when $\sigma_i$ is in the intermediate regime.

\medskip

This quantitative estimate of $\Delta{H}$ depends on the distribution in the landscape of the parameter predictions. 
Since this is unknown, no trustable estimate can be performed. However we can illustrate the dependence on the distribution, and obtain tentative results, by considering two specific limiting cases.

\begin{figure}[t]
$$\includegraphics[width=0.7\textwidth]{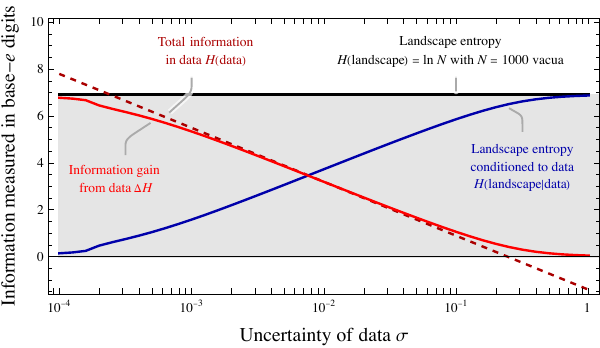}$$
\caption{\label{fig:LandscapeEntropy} We consider a toy landscape with $N=1000$ vacua,
so that the black horizontal curve at $H({\rm landscape})=\ln 1000$ shows the
Shannon landscape information entropy.
This progressively decreases down to $H({\rm landscape}|{\rm data})$
as fundamental parameters are measured
with smaller uncertainty $\sigma$ (blue curve).
The  information gain $\Delta H = H({\rm landscape})-H({\rm landscape}|{\rm data})$ from data (red curve)
agrees with the landscape-independent amount of information in data $H({\rm data})$ (dashed red curve) 
in the intermediate region where the landscape is only partially resolved.\label{plot}}
\end{figure}

First, we assume a generic dense feature-less distribution probability.
As a concrete example (results will not depend on the specific assumptions),
we assume that the $y_i$ parameters predictions are uniformly distributed in intervals $y_i\in[\mu_i-1/2,\mu_i+1/2]$, namely that the $y_{i,\va}$ predictions are equally spaced on a square lattice with lattice spacing $a=1/N^{1/n}$, around central values $\mu_i$. The critical uncertainty below which single vacua can be resolved, such that the entropy vanishes and $\Delta{H}=\ln{N}$, is $\sigma_i^c=a$. 
Uncertainties larger than the length of the $y_i$ intervals do not entail information gain and $\Delta{H}=0$. 
In the intermediate region $\sigma_i^c\ll\sigma_i\ll 1$, $\wp(y)$ in eq.~(\ref{py}) is the sum of many Gaussians with the same height, with uniformly spaced centers and a width that is much larger than the separation between the centers. Therefore it approaches the uniform distribution $\wp(y)\hspace{-2pt}\simeq \hspace{-2pt}1$ and the information gain can be approximated as
\beq\label{eq:dHu}
\Delta{H}\simeq
 \sum_{\va=1}^{N} \frac1{N}\int d^ny \,\wp(\vecf{y}\,|\va) \ln  {\wp(\vecf{y}\,|\va)}=-\sum_{i=1}^{n}\ln(\sqrt{2\pi e}\sigma_i)\,.
\eeq
The validity eq.\eq{dHu} is verified in fig.~\ref{plot} in the toy example of a landscape where $N=1000$
and $n=1$:
we see that in the intermediate region, where information partially reduces the entropy,
the generic formula of eq.\eq{dHu} reproduces the result of eq.\eq{deltaH} that takes
into account the detailed structure of the landscape.

Let us elaborate how the result depends on the unknown landscape distribution  for $y$. 
For instance if each $y_i$ was still uniformly distributed, but in an interval of length $\Delta y \neq1$, 
in eq.\eq{dHu} we would have to rescale $\sigma_i\to\sigma_i/\Delta y$, 
finding a finite correction to the universal log-enhanced term.

\smallskip

A less trivial case ois when the parameters are logarithmically distributed, namely when the logarithm of the predictions $\ln(y_{i,\va})$ are equally spaced in $[\ln(\mu_i)-1/2,\ln(\mu_i)+1/2]$ intervals. 
This situation likely has physical relevance, as the Yukawa couplings and the mass scales in the SM
have small numerical values, possibly because of unknown physical mechanisms that naturally give small
values, i.e.\ roughly logarithmic distributions. In this case the landscape probability density is $\wp(y)\simeq \prod_i 1/y_i$, in the intermediated regime for the measurement uncertainties, eq.\eq{deltaH} becomes
\beq\label{eq:dHl}
\Delta{H}\simeq
 \sum_{\va=1}^{N} \frac1{N}\int d^ny \,\wp(\vecf{y}\,|\va) \sum_{i=1}^n \ln [ y_i\, {\wp(\vecf{y}\,|\va)}]\simeq
 -\sum_{i=1}^{n}\ln(\sqrt{2\pi e}\sigma_i/\mu_i)\,.
\eeq
The second equality only holds for measurements with good relative accuracy $\sigma_i/\mu_i\ll1$.

\begin{table}[t]
$$\begin{array}{c|lc|cc} \rowcolor[rgb]{0.99,0.97,0.97}
& \hbox{Model} & \hbox{Number of} &\multicolumn{2}{c}{\hbox{Measured bits in base-$e$}}\\ 
 \rowcolor[rgb]{0.99,0.97,0.97}
\hbox{Symbol} & \hbox{description} & \hbox{parameters} & \hbox{including 0} & \hbox{without 0}\\ \hline
g_{1,2,3} & \hbox{SM gauge couplings} & 3 &37& 36\\
\lambda_H &  \hbox{SM Higgs quartic} & 1 &6& 6\\
y_q &\hbox{SM diagonal Yukawas of quarks} & 6 &50& 12 \\
y_\ell & \hbox{SM diagonal Yukawas of leptons} & 3 &72& 47\\
V_{\rm CKM} &\hbox{SM off-diagonal Yukawas of quarks} & 4 &21& 11\\
m_\nu& \hbox{Mass matrix of neutrinos} & 5 &46& 9\\
v^2/M^2_{\rm Pl}, V/M^4_{\rm Pl} & \hbox{SM/$\Lambda$CDM mass scales} &2 &371& 10\\ 
\Omega_{m,b,r}, A_s, n_s & \hbox{$\Lambda$CDM cosmological parameters} & 5 &51 & 19\\
\hline
\multicolumn{2}{c}{\hbox{All physics}}& 29 &655& 150
\end{array}$$
\caption{\label{tab:SMcoordinates}The free fundamental parameters 
of the Standard Model of particle physics and of cosmology,
that act as `coordinates' in the landscape, 
and the number of digits to which they have been measured so far.
Some digits get lost, due to QCD uncertainties, when renormalized to higher energy.
For a parameter measured as $\mu\pm \sigma$,
the column `digits including 0' gives $-\ln\sqrt{2\pi e} \sigma$, as in eq.\eq{dHu}.
The column `digits without 0' gives $-\ln\sqrt{2\pi e} \sigma/\mu$, as in eq.\eq{dHl}.
One or the other can be relevant, depending on how small numbers arise in the landscape.
We have not included the bound on $\theta_{\rm QCD}$, as the possible existence on axion would
allow to relax it.}
\end{table}%

\subsection{Practical physical discussion}
This above information-theory expressions just quantify common-sense.
For example, let us apply eq.\eq{dHu} to the top Yukawa coupling renormalized at the weak scale.
Naively, one would tell that the measurement $y_t = 0.94\pm 0.03$ provides
about one or two digits of information in base 10.
The Shannon $\Delta H$ tells that $y_t$ contains 2.1 digits of information in base $e$.
So $ \Delta H/\ln {10}$ is the number of digits in base 10, and $ \Delta H/\ln 2$ is the number of bits.

Let us next consider a fundamental physical parameter with small values,
such as the muon Yukawa coupling, $ y_\mu \approx 0.00060687$.
Including the zeros, it provides about 8 digits of information, in agreement with eq.\eq{dHu}.
Without including the zeros, it only provides about 5 digits of information, in agreement with eq.\eq{dHl}.
Indeed the zeros do not provide information that efficiently
allows to locate the SM in the landscape if the landscape contains mechanisms that produce small values
with high probability (such as a flavour symmetry for the small muon Yukawa coupling, or supersymmetry for
the small squared Higgs mass in Planck units).

In addition to experimental uncertainties, computing fundamental parameters $y_{i,v}$ from landscape vacua
will have theoretical uncertainties, that should too be included in $\sigma_i$.
Let us discuss this issue in the context of string theory.
String theory seems to need supersymmetry. 
So far only some supersymmetric string vacua have been computed, 
and only partially, finding $N=0$ vacua compatible with the SM,
and that many features of low-energy physics are left undetermined, 
as supersymmetry leaves flat directions in field space.
The experimental observation that Nature did not use supersymmetry to keep the Higgs mass fully natural
suggests, in the string landscape context, the existence of a huger number of vacua
with supersymmetry broken at the string scale.
String-theory computations have not yet clarified this issue~\cite{hep-th/0411183}.
Non-supersymmetric  vacua are presently uncomputable: we cannot even compute if they exist.
If they exist, such vacua can in principle predict values of fundamental constants,
as the lack of supersymmetric cancellations allows dynamics to generate minima rather than flat direction.

Presumably, such computations will have theoretical uncertainties such as missing higher-order terms.
Presumably, this could mean that theory will not match the high experimental accuracy in parameters such as $y_\mu$.
For sure, QCD theoretical uncertainties make part of the measured information unavailable already for the translation of the measurements into determination of the SM parameters: 
for example $\alpha_{\rm em}(M_Z)$ is much less accurately known than its low-energy value.

We proceed ignoring theoretical uncertainties on the landscape predictions and quantify the amount of information provided by measurements performed so far.
The result is shown in table~\ref{tab:SMcoordinates}.
If `0' count as information, we so far measured about 655 $e$-bits of information, reducing the landscape entropy by the same amount.
If instead small parameters are not rare in the landscape, such that `0' do not have discriminatory power,
the amount of measured information is about 160 $e$-bits.

This amount of measured information should be compared with the unknown amount of diversity in the landscape.
For example locating the SM in a landscape of $10^{500}\approx e^{1150}$ vacua roughly uniformly distributed 
would need measuring $\Delta H \sim 1150$ $e$-bits.
The landscape might have some special structure, making information related to it more relevant.
This is why in the next section we focus on measuring strategic parameters that tell if the SM has a deeper vacuum around the Planck scale.

\small

\paragraph{Acknowledgements}
We thanks Franco Cervelli, Maurizio Pierini, Philipp Roloff, Alessandro Variola, Marcel Vos, Marco Zanetti, Filip Zarnecki
for useful discussions.
This work was supported by Italian MIUR under PRIN 2017FMJFMW.




\end{document}